\documentclass[aps,psfig,epsfig,preprint] {revtex4}
\usepackage{epsfig}
\usepackage{color}
\usepackage{bm}

\begin{document}
\draft
\title{
{\normalsize \hskip4.2in USTC-ICTS-06-05} \\{\bf Nucleon-antinucleon
Interaction from the Modified Skyrme Model}}

\author{Gui-Jun Ding\footnote{e-mail address: dinggj@mail.ustc.edu.cn},
Mu-Lin Yan\footnote{e-mail address: mlyan@ustc.edu.cn}}

\affiliation{\centerline{Interdisciplinary Center for Theoretical
Study,} \centerline{University of Science and Technology of China,
Hefei, Anhui 230026, China} }

\begin{abstract}
We calculate the static nucleon-antinucleon interaction potential
from the modified Skyrme model with additional $B^{\mu}B_{\mu}$ term
using the product ansatz. The static properties of single baryon are
improved in the modified Skyrme model. State mixing is taken into
account by perturbation theory, which substantially increases the
strength of the central attraction. We obtain a long and mid range
potential which is in qualitative agreement with some
phenomenological potentials.

\end{abstract}
\maketitle
\section{introduction}
The Skyrme model is considered as the low energy limit of the
quantum chromodynamics(QCD), it models QCD in the classical or large
number of colors ($N_{C}$) limit and baryon is regarded as the
soliton in the pion field \cite{skyrme1,skyrme2,adkins1,adkins2}.
Upon quantizing a slowly rotating Skyrmion, static property of
nucleons and $\Delta$ have been calculated with the results in
agreement with experimental data within 30\% \cite{adkins1,adkins2}.
Recently it has been widely used to discuss the exotic hadrons
\cite{exotic,YLWM,DY05}. The minimal version of the model consists
of the following Lagrangian terms: the non-linear Sigma term with
chiral order $\mathcal{O}(p^2)$ and the Skyrme term with
$\mathcal{O}(p^4)$. Even though the minimal version of Skyrme  model
(Min-SKM)  can be regarded as a successful phenomenological model in
spite of its simplicity, it can not be used to study the problem of
quark spin contents of proton or EMC effects \cite{yan,yan1,Dia}
which are important QCD effects in baryon physics. This is a very
unsatisfied defect for Min-SKM. In order to cure this disease,
additional terms with $\mathcal{O}(p^6)$ or high orders  have to be
added into the model's Lagrangian to construct modified Skyrme
Models. Among them, the simplest one is the model with the Min-SKM
Lagrangian plus only one additional $B_\mu B^\mu$ term \cite{yan1},
where $B_\mu$ is the baryon current ( or Goldstone-Wilczek current
). Hereafter, we shortly call this simplest modified Skyrme model as
Mod-SKM. It is expected that the Mod-SKM should be more realistic
than Min-SKM. To discuss this issue, and to fix the parameters in
Mod-SKM is one of the aims of this paper. Moreover the Mod-SKM can
be obtained by considering the infinite $\omega$ mass limit of the
vector meson $\omega$ term of the chiral Lagrangian studied in Ref
\cite{adkins3}.

An interesting application of the Skyrme model is the investigation
of the baryon-baryon interaction, especially the
nucleon-nucleon($NN$) interaction
\cite{nucleon1,nucleon2,nucleon3,nucleon4,nucleon5,nucleon6}. The
Skyrme picture gives us a qualitative understanding of the principal
features of the $NN$ interaction: it has the correct long-range one
pion exchange potential which dominates the tensor force. There is a
strong short range repulsion, and finally there is a pronounced
central attraction at intermediate range, albeit weakly attractive
while comparing with the phenomenological potential. However, the
recent development of obtaining the $NN$ interaction from the Skyrme
model has shown that the combined effect of the careful treatment of
the nonlinear equations and the configuration mixing is to give
substantial central midrange attraction for the $NN$ system that is
in qualitative agreement with the data \cite{nucleon5}.

The $NN$ and $N\overline{N}$ potentials have been investigated by
means of the Min-SKM and the algebraic methods in Ref.
\cite{amado1,amado2,amado3}. The phenomenon and puzzles in the
baryon-antibaryon physics have attracted much attention recently due
to the remarkable discovery of baryon-antibaryon enhancements in the
$J/\psi$ and $B$ decays \cite{Bes1,Bes2,Bes3,belle1,belle2,belle3}.
The $N\overline{N}$ interaction and the possible nucleon-antinucleon
bound states have been investigated from the constituent quark model
\cite{constituent,DY06,rDY06}. In the Skyrme model, the interactions
between  classical Skyrmion and antiskyrmion, i.e., $S\overline{S}$,
were explored in  \cite{YLWM,DY05}. In the present paper, we shall
study the $N \overline{N}$ potential using the Mod-SKM and following
the methods developed in Refs. \cite{nucleon5,amado1,amado2}.

It is well-known that phenomenologically the $N\overline{N}$
potential is not as well established as the $NN$ potential. At
distance less than about 1 fermi the interaction is dominated by
annihilation. However, at larger distance, a meaningful potential
can be defined and studied either by $G$ parity transformation on
the $NN$ meson exchange potential or phenomenologically.

We will compare our Mod-SKM's results to some phenomenological
potentials. The $B^{\mu}B_{\mu}$ term in Mod-SKM reflects the effect
of $\omega$ meson exchange \cite{adkins3,jackson,oka}. We will see
that at large distance, where the product ansatz makes the best
sense, the potentials based on the Skyrme model agree qualitatively
and, in most cases, quantitatively with the phenomenological
interactions. At intermediate and short distance, we do less well,
but at these distances the product ansatz is not valid. However, the
results are still suggestive.

In the following section, we give the Mod-SKM's Lagrangian, then
reproduce a number of static properties of single baryon which is
both qualitatively appealing and quantitatively satisfactory. In
Sec.III we study the skyrmion-antiskyrmion interaction in Mod-SKM,
and project them to the nucleon space by the algebraic methods
\cite{amado1,amado2,amado3}. We also consider the effects of
rotational excitations by including intermediate states $\Delta$ and
$\overline{\Delta}$, and evaluate the corrections to the
$N\overline{N}$ potential in perturbation theory. Sec.IV closes this
paper with some discussions related to the present study.

\section{the modified skyrme model and the static properties of single baryon}
The Skyrme model lagrangian is generalized to include additional
$B^{\mu}B_{\mu}$ term which simulates the effects of $\omega$ meson,
and this modified Skyrme model lagrangian provides a better
description of both the single baryon static properties and the low
energy $NN$ interaction \cite{jackson,oka}. This lagrangian has the
following form
\begin{equation}
\label{1}{\cal
L}=\frac{F^2_{\pi}}{16}{\rm{Tr}}(\partial_{\mu}U\partial^{\mu}U^{\dag})+\frac{1}{32e^2}
{\rm{Tr}}([U^{\dagger}\partial_{\mu}U,U^{\dagger}\partial_{\nu}U]^2)+\frac{1}{8}m^2_{\pi}F^2_{\pi}{\rm{Tr}}(U-1)
-\frac{3\pi^2N_C}{5m^2}B^{\mu}B_{\mu}
\end{equation}
where $U$ is a $SU(2)$ valued field $U=\exp[2i\tau_a\pi_a/F_{\pi}]$,
$B^{\mu}$ is the topological current
$B^{\mu}=\frac{1}{24\pi^2}\varepsilon^{\mu\nu\alpha\beta}{\rm{Tr}}[(U^{\dagger}\partial_{\nu}U)
(U^{\dagger}\partial_{\alpha}U)(U^{\dagger}\partial_{\beta}U)]$, and
$e,\;F_{\pi},\; m$ are parameters to be determined. The first term
is the lagrangian of meson fields in the nonlinear sigma model and
the second term is the so called Skyrme term which stabilizes the
soliton. The third term is the pion mass term and the fourth term is
the additional $B^{\mu}B_{\mu}$ term. $U$ transforms as
$U\rightarrow U'=LUR^{\dagger}$ under the chiral group
$SU(2)_{L}\times SU(2)_{R}$, where both $L$ and $R$ are $SU(2)$
matrices.

The chiral soliton model \cite{adkins3} where, as an alternative to
the Skyrme term, the vector meson $\omega$ term $\beta
\omega^{\mu}B_\mu$ stabilizes the soliton, provides a support for
the interpretation of the $B_{\mu}B^{\mu}$ term which emerges in the
limit $m_{\omega}\rightarrow\infty$. In traditional nuclear
interaction theories within the potential framework which are based
on the single meson exchange mainly, it is shown that
$N\overline{N}$ system is more attractive than the $NN$ system due
to the fact that in the theories there is a strong $\omega$
exchange, so inclusion of this term which models the effect of the
$\omega$ meson could help in a better description of the  $N \bar N$
system. Furthermore, the study of the quark spin content also
support that we should add this six derivative term in order to
yield a spin content consistent with the present experiment
\cite{yan,spin}. Generally terms in ${\cal L}$ with more than two
time derivatives can lead to pathological runaway solutions when the
adiabatic approximation is relaxed and present obvious difficulties
in quantizing the theory. But the Lagrangian of Mod-SKM have, at
most, two time derivatives, hence there is no such difficulty.

For the case with single static Skyrmion, we use the so called
hedgehog ansatz:
\begin{equation}
\label{2}U_0(\mathbf{r})=\exp[i\tau_a\hat{r}_a F(r)]
\end{equation}
where $F(r)$ is the chiral angle which minimizes the static soliton
energy subject to the boundary condition $F(0)=\pi$~and~
$F(\infty)=0$. From Eq. (\ref{1})and Eq. (\ref{2}), we get the mass
of classical soliton:
\begin{equation}
\label{3}M_s=\frac{\pi}{2}\frac{F_{\pi}}{e}\int_{0}^{\infty}\{x^2F'^2+2S^2+4S^2(2F'^2+\frac{S^2}{x^2})+
2\mu^2x^2(1-C)+\nu^2\frac{S^4}{x^2}F'^2\}
\end{equation}
with
\begin{eqnarray}
\nonumber && x=eF_{\pi}r,~~~
\mu^{2}=\frac{m^{2}_{\pi}}{e^2F^2_{\pi}}\;,
~~~F'=\frac{dF}{dx}\;,\\
\label{4}&&\nu^2=\frac{18e^4F^2_{\pi}}{5\pi^2m^2}\;,~~~
S={\sin}F\;,~~~ C={\cos}F.
\end{eqnarray}
In Eq. (\ref{3}), the term proportional to $\nu^2$  comes from the
$B^\mu B_\mu $ term and is absent in the conventional Skyrme model.
Minimizing $M_s$ with respect to $F$, $\delta M_s=0$, we have the
following equation for $F$,
\begin{equation}
\label{5}(\frac{x^2}{4}+2S^2+\frac{\nu^2}{4}\frac{S^4}{x^2})F''+(\frac{\nu^2}{2}\frac{S^3C}{x^2}+2SC)F'^2
+(\frac{x}{2}-\frac{\nu^2}{2}\frac{S^4}{x^3})F'-(\frac{1}{2}SC+\frac{2S^3C}{x^2}+\frac{1}{4}\mu^2x^2S)=0
\end{equation}
From the above equation, we can see that the chiral angle $F$
asymptotically tends to the following expression when $r$ goes to
infinity,
\begin{equation}
\label{6}F(r)\rightarrow
\mathcal{A}(\frac{1}{m_{\pi}eF_{\pi}r^2}+\frac{1}{eF_{\pi}r})\;e^{-m_{\pi}r}
~~~(r\rightarrow\infty)
\end{equation}
The coefficient $A$ is related to the pion-nucleon coupling constant
$g_{{\pi}NN}$ through \cite{adkins1},
\begin{equation}
\label{7}g_{{\pi}NN}=\frac{4{\pi}M_{N}\mathcal{A}}{3e m_{\pi}}
\end{equation}
Associated with the chiral symmetry, the vector current
$J^{{\mu}a}_{V}$ and axial vector current $J^{{\mu}a}_{A}$ can be
obtained from the Skyrme lagrangian Eq. (\ref{1}) following the
standard procedure,
\begin{eqnarray}
\nonumber&&J^{{\mu}a}_{V}=\frac{iF^2_{\pi}}{8}{\rm{Tr}}[\frac{\tau_{a}}{2}\;(\partial^{\mu}UU^{\dag}+\partial^{\mu}U^{\dag}U)]
-\frac{i}{8e^2}{\rm{Tr}}\{[\frac{\tau_a}{2},\partial_{\nu}UU^{\dagger}][\partial^{\mu}UU^{\dagger},\partial^{\nu}UU^{\dagger}]\\
\label{8}&&+[\frac{\tau_a}{2},\partial_{\nu}U^{\dagger}U][\partial^{\mu}U^{\dagger}U,\partial^{\nu}U^{\dagger}U]\}-\frac{3N_{C}i}{20m^2}
\;\varepsilon^{\mu\nu\alpha\beta}B_{\nu}{\rm{Tr}}[\frac{\tau_{a}}{2}(\partial_{\alpha}UU^{\dag}\partial_{\beta}UU^{\dag}-\partial_{\alpha}U^{\dag}U\partial_{\beta}U^{\dag}U)],\\
\nonumber&&J^{{\mu}a}_{A}=\frac{iF^2_{\pi}}{8}{\rm{Tr}}[\frac{\tau_{a}}{2}\;(\partial^{\mu}UU^{\dag}-\partial^{\mu}U^{\dag}U)]
-\frac{i}{8e^2}{\rm{Tr}}\{[\frac{\tau_a}{2},\partial_{\nu}UU^{\dagger}][\partial^{\mu}UU^{\dagger},\partial^{\nu}UU^{\dagger}]\\
\label{9}&&-[\frac{\tau_a}{2},\partial_{\nu}U^{\dagger}U][\partial^{\mu}U^{\dagger}U,\partial^{\nu}U^{\dagger}U]\}-\frac{3N_{C}i}{20m^2}
\;\varepsilon^{\mu\nu\alpha\beta}B_{\nu}{\rm{Tr}}[\frac{\tau_{a}}{2}(\partial_{\alpha}UU^{\dag}\partial_{\beta}UU^{\dag}+\partial_{\alpha}U^{\dag}U\partial_{\beta}U^{\dag}U)]
\end{eqnarray}
The classical field configuration of the hedgehog form does not have
definite spin and isospin. However, nucleons carry both spin and
isospin, and in any reasonable model of nucleons the appropriate
spin and isospin states must appear. Following the conventional way,
we perform the collective coordinate quantization. We make a time
dependent $SU(2)$ rotation of our static soliton solution,
\begin{equation}
\label{10}U(x)=A(t)U_0(x)A^{\dag}(t)
\end{equation}
then
\begin{equation}
\label{11}L=-M_s+I\;\rm{Tr}(\partial_0A^{\dagger}(t)\partial_0A(t))
\end{equation}
where $A(t)\in SU(2)$-matrix is the collective coordinate, and $I$
is the moment of inertia, which is given by
\begin{equation}
\label{12}I=\frac{1}{F_{\pi}e^3}\frac{2\pi}{3}\int^{\infty}_{0}dx\{S^2[x^2+4(x^2F'^2+S^2)]+\nu^2S^4F'^2\}
\end{equation}
If the $SU(2)$ matrix $A(t)$ is parameterized by
$A(t)=a_0+i\tau_{n}a_n$, with $a^2_0+\sum^{3}_{n=1}a^2_n=1$, the
Hamiltonian is
\begin{equation}
\label{13}H=M_s-\frac{1}{8I}\sum^{3}_{n=0}(\frac{\partial}{{\partial}a_n})^2=M_s+\frac{\mathbf{S}^2}{2I}=M_s+\frac{\mathbf{I}^2}{2I}
\end{equation}
Noting that the $I$ in the denominator is the moment of inertia,
$\mathbf{S}$ and $\mathbf{I}$ are  the spin and isospin operators
respectively. As in Ref. \cite{adkins1}, we can calculate the static
properties of single baryon. In going from the classical results to
the quantum results for rotation operators we must symmetrize them
\cite{spin}, {\it i.e.,} we perform the Weyl order of these
operators.

From Eq. (\ref{13}) the masses of nucleon and $\Delta$ respectively
are
\begin{equation}
\label{14}M_N=M_s+\frac{3}{8I}\;,~~~~~M_{\Delta}=M_s+\frac{15}{8I}
\end{equation}
The isoscalar and isovector mean square electric radii are
\begin{eqnarray}
\label{15}&&\langle
r^2\rangle_{E,I=0}=\frac{1}{(eF_{\pi})^2}\int^{\infty}_{0}dx\frac{-2}{\pi}x^2S^2F'\\
\label{16}&&\langle
r^2\rangle_{E,I=1}=\frac{1}{(eF_{\pi})^2}\frac{1}{Ie^3F_{\pi}}\int^{\infty}_{0}dx\{\frac{2\pi}{3}
x^4S^2[1+4(F'^2+\frac{S^2}{x^2})]+\frac{2\pi}{3}\nu^2x^2S^4F'^2\}
\end{eqnarray}
The corresponding proton and neutron mean square charge radii are
\begin{eqnarray}
\nonumber&&\langle r^2\rangle_{E,p}=\frac{1}{2}(\langle
r^2\rangle_{E,I=0}+\langle r^2\rangle_{E,I=1})\;,\\
\label{17}&&\langle r^2\rangle_{E,n}=\frac{1}{2}(\langle
r^2\rangle_{E,I=0}-\langle r^2\rangle_{E,I=1})
\end{eqnarray}
After some somewhat tedious but straightforward calculations, we can
obtain the proton and neutron's magnetic moment which are
respectively,
\begin{eqnarray}
\nonumber&&\mu_{p}=2M_N(\frac{1}{12I}\langle
r^2\rangle_{E,I=0}+\frac{I}{6})\;,\\
\label{18}&&\mu_{n}=2M_N(\frac{1}{12I}\langle
r^2\rangle_{E,I=0}-\frac{I}{6})
\end{eqnarray}
In  the above, we have symmetrized the rotation operators, and the
proton and neutron's magnetic momentum are defined through
\begin{eqnarray}
\nonumber&&\langle p,1/2|\mu^3|p,1/2\rangle=\frac{1}{2M_N}\mu_p\;,\\
\label{19}&&\langle n,1/2|\mu^3|n,1/2\rangle=\frac{1}{2M_N}\mu_n
\end{eqnarray}
After some lengthy calculations, we can also get the axial coupling
constant\cite{adkins1}:
\begin{eqnarray}
\nonumber
g_A&=&-\frac{2\pi}{9e^2}\int^{\infty}_{0}dxx^2\{\frac{2CS}{x}[1+4(F'^2+\frac{S^2}{x^2})]+F'(1+\frac{8S^2}{x^2})\}-\frac{12}{5\pi
m^2}\int^{\infty}_{0}dxS^2F'\{(eF_{\pi})^2(\frac{2CS}{3x}F'\\
\label{20}&&+\frac{S^2}{3x^2})-(4xCSF'+S^2)\frac{1}{18I^2}\}
\end{eqnarray}
There are three parameters in the modified Skyrme model, {\it i.e.,}
$e$, $F_{\pi}$, $m$, the pion mass is $m_{\pi}=138\rm{MeV}$, and the
number of color $N_C=3$. In the conventional Skyrme model, there is
always a conflict between  $e$- and $F_{\pi}$-datum input setting
for giving correct nucleon and $\Delta$ masses or  giving correct
strength of the pion tail \cite{nucleon3,nucleon5,amado2}. But a
satisfactorily simultaneous description of the nucleon, $\Delta$
mass and the strength of the pion tail is possible by properly
choosing the parameters $e$,$F_{\pi}$, $m$ in Mod-SKM. Throughout
our calculation we choose the three parameters as
$e=19.48,\;\;F_{\pi}=129.11\rm{MeV},\;\;m=420\rm{MeV}$, this
parameter setting gives $g_{{\pi}NN}\approx13.5$ through Eq.
(\ref{7}), which leads to the correct one-pion exchange potential of
$N\overline{N}$ interaction as the distance tends to infinity. The
connection between the Mod-SKM and the chiral soliton model
including $\omega$ meson \cite{adkins3}, allows us relate the
parameter $m$ to the coupling $\beta$,  i.e.,
$m=\sqrt{\frac{2}{5}}\frac{3\pi m_{\omega}}{\beta}$, and the best
fit of the parameters in Ref. \cite{adkins3} gives
$m\approx298.8$MeV, which is not too far from the value of $m$ in
this work. The static properties of single baryon are summarized in
the Table I, and the results of conventional Skyrme model
\cite{adkins2} as well as the experimental values are also given in
Table I.

\begin{table}[hptb]
\begin{center}
\caption{Static properties of single baryon in the modified Skyrme
model compared to those in the conventional one \cite{adkins2} and
to the experimental results.}
\begin{tabular}{|c|c|c|c|}\hline\hline
Physical Quantity & Modified Skyrme Model & Conventional Skyrme
Model & Experiment Results\\ \hline
$M_N$&938.9 \rm{MeV}(input)&938.9 \rm{MeV}(input)&938.9 \rm{MeV}\\
\hline
$M_{\Delta}$&1232 \rm{MeV}(input)&1232\ rm{MeV}(input)&1232 \rm{MeV}\\
\hline
$M_{\pi}$&138 \rm{MeV}(input)&138 \rm{MeV}(input)&138 \rm{MeV}\\
\hline
$e$&19.48&4.84&---\\
\hline
$F_{\pi}$&129.11 \rm{MeV}&108 \rm{MeV}&186 \rm{MeV}\\
\hline
$\langle r^2\rangle^{1/2}_{E,I=0}$&0.71 \rm{fm}&0.68 \rm{fm}&0.72 \rm{fm}\\
\hline
$\langle r^2\rangle^{1/2}_{E,I=1}$&1.04 \rm{fm}&1.04 \rm{fm}&0.88 \rm{fm}\\
\hline
$\mu_{p}$&2.01&1.97&2.79\\
\hline
$\mu_{n}$&-1.20&-1.23&-1.91\\
\hline
$g_{A}$&0.82&0.65&1.24\\
\hline\hline
\end{tabular}
\end{center}
\end{table}

In Table I we can see that the prediction  of the Mod-SKM are closer
to the experimental values than  those of the conventional Skyrme
model \cite{adkins2}, so we expect that Mod-SKM provides a better
description to other static properties of baryons including the low
energy $N\overline{N}$ interaction.

\section{adiabatic $N\overline{N}$ interaction}

\subsection{FORMULATION}

We now study, in the product ansatz, the interaction energy of the
Skyrmion-antiSkyrmion($S\overline{S}$) system, which is
 a function of separation between $S$ and $\overline{S}$ and the relative orientation.
This interaction energy  can be calculated numerically. We rotate
the two solitons independently in $SU(2)$ space,
\begin{eqnarray}
\nonumber&&U_0(\mathbf{r}-\mathbf{R}/2)\rightarrow
AU_0(\mathbf{r}-\mathbf{R}/2)A^{\dagger}\;,\\
\label{21}&&U^{\dagger}_0(\mathbf{r}+\mathbf{R}/2)\rightarrow
BU^{\dagger}_0(\mathbf{r}+\mathbf{R}/2)B^{\dagger}
\end{eqnarray}
where both $A$ and $B$ are $SU(2)$ matrices. In order to obtain the
static $N\overline{N}$ interaction, we describe the $N\overline{N}$
configuration with the product ansatz (exact in the large $R$ limit)
as follows,
\begin{equation}
\label{22}
U(\mathbf{r})=AU_0(\mathbf{r}-\mathbf{R}/2)A^{\dagger}BU^{\dagger}_0(\mathbf{r}+\mathbf{R}/2)B^{\dagger},
\end{equation}
where one baryon located at $\mathbf{R}/2$ and the antibaryon at
$-\mathbf{R}/2$. Retaining only the potential energy density in the
modified Skyrme lagrangian (\ref{1}), the energy in the field of Eq.
(\ref{22}) is the same as in
\begin{equation}
\label{23}
U(\mathbf{r})=U_0(\mathbf{r}-\mathbf{R}/2)CU^{\dagger}_0(\mathbf{r}+\mathbf{R}/2)C^{\dagger}
\end{equation}
where $C=A^{\dagger}B=c_4+i\bm{\tau}\cdot\mathbf{c}$ is a $SU(2)$
matrix too. Discarding non-static terms containing time derivatives,
the static $N\overline{N}$ potential is defined by,
\begin{equation}
\label{24}V(\mathbf{R},C)=-\int d^3x{\cal L}[U(\mathbf{r})]-2M_s
\end{equation}
$V(\mathbf{R},C)$ can be written in the notation of Vinh Mau {\it et
al.} \cite{nucleon2} as
\begin{equation}
\label{25}V(\mathbf{R},C)=V_1(R)+V_2(R)c^2_4+V_3(R)(\mathbf{c}\cdot\mathbf{\hat{R}})^2+V_4(R)c^4_4+V_5(R)c^2_4(\mathbf{c}\cdot\mathbf{\hat{R}})^2
+V_6(R)(\mathbf{c}\cdot\mathbf{\hat{R}})^4
\end{equation}
where $V_i(i=1-6)$ are functions of $R$. Generally, for
$S\overline{S}$, the symmetry $\mathbf{R}\rightarrow-\mathbf{R}$ is
broken by the product ansatz, and we need three additional terms for
a consistent expansion,
\begin{eqnarray}
\nonumber
V(\mathbf{R},C)&=&V_1(R)+V_2(R)c^2_4+V_3(R)(\mathbf{c}\cdot\mathbf{\hat{R}})^2+V_4(R)c^4_4+V_5(R)c^2_4(\mathbf{c}\cdot\mathbf{\hat{R}})^2
+V_6(R)(\mathbf{c}\cdot\mathbf{\hat{R}})^4\\
\label{26}&&+V_7(R)c_4(\mathbf{c}\cdot\mathbf{\hat{R}})+V_8(R)c^3_4(\mathbf{c}\cdot\mathbf{\hat{R}})+V_9(R)c_4(\mathbf{c}\cdot\mathbf{\hat{R}})^3
\end{eqnarray}
These terms odd in $\mathbf{R}$ are artifacts of the symmetry of the
product ansatz and should be discarded. One can use the symmetrized
energy $\frac{V(\mathbf{R},C)+V(-\mathbf{R},C)}{2}$ to extract
$V_1(R)$ to $V_6(R)$, since the $V_7(R)$ to $V_9(R)$ terms drop out
in this combination.

Next, we have to map the Skyrmion-antiSkyrmion($S\overline{S}$)
interaction to the nucleon-antinicleon($N\overline{N}$) interaction.
This problem has been tackled in various ways by various groups for
the $NN$ case \cite{nucleon1,nucleon2,amado1}. Each of the forms
used in these works can always be cast in the form of the algebraic
model \cite{amado1}. So we will also use the algebraic method for
mapping the $S\overline{S}$ interaction to the $N\overline{N}$
interaction \cite{nucleon5,amado1,amado2}. This method allows us to
study both the large $N_C$ limit, as well as to include the finite
$N_C$ effects explicitly in a systematic way. Most of the formulas
given below can be found in Refs. \cite{nucleon5,amado1,amado2},
however for the sake of completeness we remind here the important
ones.

The algebraic model  consists of two sets of $U(4)$ algebras, one
for each Skyrmion (or antiSkyrmion), as well as the radial
coordinate $\mathbf{R}$. This method was developed in Ref.
\cite{nucleon5,amado1} for the $NN$ system, and also generalized to
the $N\overline{N}$ system \cite{amado2,amado3}. In large $N_C$
limit, the $S\overline{S}$ interaction can be expanded in terms of
three operators: the identity, the operator $W$ and the operator
$Z$,
\begin{eqnarray}
\nonumber&&W=T^{\alpha}_{pi}T^{\beta}_{pi}/N^2_C\;,\\
\label{27}&&Z=T^{\alpha}_{pi}T^{\beta}_{pj}[3\hat{R}_{i}\hat{R}_{j}-\delta_{ij}]/N^2_C.
\end{eqnarray}
Here $\alpha$ and $\beta$ label two different sets of bosons, used
to realize the $U(4)$ algebra, and $T$ is an one-body operator with
spin and isospin $\mathbf{1}$. The semiclassical (large $N_C$) limit
of these operators can be given in terms of $\mathbf{\hat{R}}$ and
$C=c_4+i\bm{\tau}\cdot\mathbf{c}$ as \cite{amado1}
\begin{eqnarray}
\nonumber W_{cl}(A,B)&=&3c^2_4-\mathbf{c}^2=4c^2_4-1\;,\\
\label{28}
Z_{cl}(A,B,\mathbf{\hat{R}})&=&6(\mathbf{c}\cdot\mathbf{\hat{R}})^2-2\mathbf{c}^2=2c^2_4-2+6(\mathbf{c}\cdot\mathbf{\hat{R}})^2
\end{eqnarray}
The $S\overline{S}$ interaction can be expressed as
\begin{equation}
\label{29}V(\mathbf{R},C)=\upsilon_1(R)+\upsilon_2(R)W_{cl}+\upsilon_3(R)Z_{cl}+\upsilon_4(R)W_{cl}^2+\upsilon_5(R)W_{cl}Z_{cl}+\upsilon_6(R)Z_{cl}^2
\end{equation}
in the semiclassical limit. We can obtain the relations between
$V_i$ and $\upsilon_i(i=1-6)$ by comparing Eq. (\ref{25}) and Eq.
(\ref{29})
\begin{eqnarray}
\nonumber&&V_1(R)=\upsilon_1(R)-\upsilon_2(R)-2\upsilon_3(R)+\upsilon_4(R)+2\upsilon_5(R)+4\upsilon_6(R)\;,\\
\nonumber&&V_2(R)=4\upsilon_2(R)+2\upsilon_3(R)-8\upsilon_4(R)-10\upsilon_5(R)-8\upsilon_6(R)\;,\\
\nonumber&&V_3(R)=6\upsilon_3(R)-6\upsilon_5(R)-24\upsilon_6(R)\;,\\
\nonumber&&V_4(R)=16\upsilon_4(R)+8\upsilon_5(R)+4\upsilon_6(R)\;,\\
\nonumber&&V_5(R)=24\upsilon_5(R)+24\upsilon_6(R)\;,\\
\label{30}&&V_6(R)=36\upsilon_6(R)
\end{eqnarray}
Six independent choices of the matrix $C$ can yield enough
independent linear equations to determine $\upsilon_i(R)(i=1-6)$ or
equivalently $V_i(R)(i=1-6)$ through Eq. (\ref{30}), and the
numerical results for $\upsilon_i(R)(i=1-6)$ coming from the Mod-SKM
are shown in Fig. 1. From this figure it can be seen that the first
three term $\upsilon_1(R)$, $\upsilon_2(R)$ and $\upsilon_3(R)$ are
dominant. It seems a good approximation to neglect the interaction
terms which are nonlinear in the expansion of operator $W$ and $Z$.
In the following discussion, we will mostly concentrate on the first
three terms, then the leading term in this expansion is given by the
following form:
\begin{equation}
\label{31}V(\mathbf{R},C)=\upsilon_1(R)+\upsilon_2(R)W+\upsilon_3(R)Z
\end{equation}

The algebraic operators $W$ and $Z$ have simple expectation values
for the nucleons \cite{amado1}
\begin{eqnarray}
\nonumber\langle
N|T^{\alpha}_{pi}|N\rangle&=&-\frac{N_{C}}{3}P_{N}\langle
N|\tau^{\alpha}_p\sigma^{\alpha}_i|N\rangle\;,\\
\nonumber\langle
N\overline{N}|W|N\overline{N}\rangle&=&\frac{1}{9}P^2_N\langle
N\overline{N}|{\bm{\sigma}^1\cdot\bm{\sigma}^2\bm{\tau}^1\cdot\bm{\tau}^2}|N\overline{N}\rangle\;,\\
\label{32}\langle
N\overline{N}|Z|N\overline{N}\rangle&=&\frac{1}{9}P^2_N\langle
N\overline{N}|(3\bm{\sigma}^1\cdot\mathbf{\hat{R}}\bm{\sigma}^2\cdot\mathbf{\hat{R}}-\bm{\sigma}^1\cdot\bm{\sigma}^2)\bm{\tau}^1\cdot\bm{\tau}^2|N\overline{N}\rangle
\end{eqnarray}
Here $P_N$ is the finite $N_C$ correction factor
$P_N=1+\frac{2}{N_C}$. By using Eq. (\ref{32}) we take the
$N\overline{N}$ matrix element of the interaction and evaluate the
$N\overline{N}$ potential, which only contains three independent
multipole component, {\it i.e.,} the central part $V_c$, the
spin-spin part $V_s$, and the tensor term $V_t$:
\begin{equation}
\label{33}V^{(0)}(\mathbf{R})=V_c(R)+V_s(R)\bm{\sigma}^1\cdot\bm{\sigma}^2\bm{\tau}^1\cdot\bm{\tau}^2+V_t(R)
(3\bm{\sigma}^1\cdot\mathbf{\hat{R}}\bm{\sigma}^2\cdot\mathbf{\hat{R}}-\bm{\sigma}^1\cdot\bm{\sigma}^2)\bm{\tau}^1\cdot\bm{\tau}^2
\end{equation}
with
\begin{equation}
\label{34}V_c=\upsilon_1\;,~~~V_s=\frac{\upsilon_2P^2_N}{9}\;,~~~V_t=\frac{\upsilon_3P^2_N}{9}
\end{equation}
The $N\overline{N}$ potential in the above is calculated by
projecting Eq. (\ref{31}) to the nucleon degrees of freedom only,
and this is the correct procedure for large separation. However, at
short distance the nucleons may deform or excite as they interact,
and they can be virtually whatever the dynamics requires, for
example, $\Delta(\rm{or}\;\overline{\Delta})$. This means that we
need to consider the state mixing effect. In the case of $NN$
interaction, we saw that states mixing plays an important role in
obtaining the phenomenologically correct potential. We expect the
state mixing effect to be very important in the $N\overline{N}$
interaction as well. The state mixing comes into effect at the
distance where the product ansatz makes no longer sense, so our
results at short and intermediate distances should be suggestive,
although we include state mixing. As a guide, we study the effects
of the intermediate states $N\overline{\Delta}$,
$\Delta\overline{N}$ and $\Delta\overline{\Delta}$ perturbatively,
then to second order, the $N\overline{N}$ interaction is given by
\begin{equation}
\label{35}V(\mathbf{R})=\langle
N\overline{N}|V(\mathbf{R},C)|N\overline{N}\rangle+\sum_s\;'\;\frac{\langle
N\overline{N}|V(\mathbf{R},C)|s\rangle\langle
s|V(\mathbf{R},C)|N\overline{N}\rangle}{E_{N\overline{N}}-E_s}
\end{equation}
Here $E_{N\overline{N}}$ is the two nucleon energy and $E_s$ is the
energy of the relevant excited state. The first term on the right is
the direct nucleon-antinucleon projection of $V(\mathbf{R},C)$ and
it is exactly the expression $V^{(0)}(\mathbf{R})$. The second term
is the correction due to rotational or excited states. It is clear
from the energy denominator that the second term is attractive. We
need to evaluate the three sets of matrix elements $\langle
N\overline{N}|V(\mathbf{R},C)|N\overline{\Delta}\rangle\langle
N\overline{\Delta}|V(\mathbf{R},C)|N\overline{N}\rangle$, $\langle
N\overline{N}|V(\mathbf{R},C)|\Delta\overline{N}\rangle\langle
\Delta\overline{N}|V(\mathbf{R},C)|N\overline{N}\rangle$ and
$\langle
N\overline{N}|V(\mathbf{R},C)|\Delta\overline{\Delta}\rangle\langle
\Delta\overline{\Delta}|V(\mathbf{R},C)|N\overline{N}\rangle$, and
the final result for the first order correction to the
$N\overline{N}$ interaction is \cite{nucleon5,amado2}
\begin{eqnarray}
\nonumber
&&V^{(1)}_{PT}(\mathbf{R})=-\frac{Q^2_N}{\delta}\{[\frac{1}{3}Q^2_NP^{\tau}_0+
(\frac{16}{27}P^2_N+\frac{5}{27}Q^2_N)P^{\tau}_1][\upsilon^2_2(R)+2\upsilon^2_3(R)]\\
\nonumber&&+(\bm{\sigma}^1\cdot\bm{\sigma}^2)[-\frac{1}{18}Q^2_NP^{\tau}_0+(\frac{16}{81}P^2_N-\frac{5}{162}Q^2_N)P^{\tau}_1][\upsilon^2_2(R)-\upsilon^2_3(R)]+
(3\bm{\sigma}^1\cdot\mathbf{\hat{R}}\bm{\sigma}^2\cdot\mathbf{\hat{R}}-\bm{\sigma}^1\cdot\bm{\sigma}^2)\\
\label{36}&&\times[-\frac{1}{18}Q^2_NP^{\tau}_0+(\frac{16}{81}P^2_N-\frac{5}{162}Q^2_N)P^{\tau}_1][\upsilon^2_3(R)-\upsilon_2(R)\upsilon_3(R)]\}
\end{eqnarray}
Here $Q_N$ is another finite $N_C$ correction factor
$Q_N=\sqrt{(1-1/N_C)(1+5/N_C)}$. $\delta$ is the $N-\Delta$ energy
difference, which is about 300 MeV, and $P^{\tau}_{T}(T=0,1)$ is a
projection operator onto the isospin $T$,
$P^{\tau}_{0}=\frac{1}{4}(1-\bm{\tau}_1\cdot\bm{\tau_2})$,
$P^{\tau}_{1}=\frac{1}{4}(3+\bm{\tau}_1\cdot\bm{\tau_2})$.

\subsection{RESULTS}
For each total isospin $T=0,1$ we parameterize the $N\overline{N}$
interaction potential by:
\begin{equation}
\label{37}V^{T}_{N\overline{N}}=V^{T}_c+V^{T}_s\bm{\sigma}^1\cdot\bm{\sigma}^2+V^{T}_{t}(3\bm{\sigma}^1\cdot\hat{\mathbf{R}}\bm{\sigma}^2\cdot\hat{\mathbf{R}}-\bm{\sigma}^1\cdot\bm{\sigma}^2)
\end{equation}
We now calculate $V^{T}_c$, $V^{T}_s$, $V^{T}_t$ for each isospin
$T$($T$=0,1) following the methods outlined above. Such a
calculation requires considerable computing time. We would like to
compare the Skyrmion model potentials with the realistic
nucleon-antinucleon interaction potentials. However, we can not
relate our results to the modern $N\overline{N}$ interaction
potential such as the Paris potential \cite{paris} and the Julich
potential \cite{julich}, since their central parts contain explicit
momentum dependent terms. For that reason we compare our results
with the phenomenological potentials of Bryan-Phillips \cite{bp} and
of the Nijmegen group \cite{nij}. These potentials provide
successful descriptions of both the $N\overline{N}$ scattering
experiments data and the spectrum of resonances, and they are not
qualitatively different from each other. At large distance all these
potentials can be correctly described by the one-boson exchange
mechanism and the $N\overline{N}$ potential can be obtained by
$G$-parity transformation of the corresponding parts of the $NN$
interaction potential. Using equation of motion and the asymptotic
form Eq. (\ref{6}) of the chiral angle $F(r)$, the $N\overline{N}$
interaction based on the Mod-SKM tends to one pion exchange
potential in the large distance region \cite{nucleon6},
\begin{equation}
\label{38}V^{N\overline{N}}(r)\rightarrow\frac{-1}{4\pi}(\frac{g_{\pi
NN}}{2M_N})^2(\bm{\tau}^1\cdot\bm{\tau}^2)(\bm{\sigma}^1\cdot\bm{\nabla})(\bm{\sigma}^2\cdot\bm{\nabla})\frac{e^{-m_{\pi}r}}{r}~~~~(r\rightarrow\infty)
\end{equation}
The parameters $e, F_{\pi}, m$ are properly chose to guarantee that
the long distance tail of the $N\overline{N}$ interaction will agree
with the phenomenology. In order to model the annihilation effect at
short distance, various cut off has been used in the Bryan-Phillips,
Nijmegen, and other similar potentials. At short distance, the
interaction is dominated by the strong absorptive potential of order
1 GeV, and it is significantly different from the meson exchange
potential. Furthermore, the Skyrme model at short distance is no
longer meaningful. so we should not take seriously the comparison of
our results with the phenomenological potentials at 1 fm and less,
however the results are still indicative at short distance. We find
that the principal feature of the phenomenological $N\overline{N}$
interaction emerges from the careful calculation of that interaction
based on the Mod-SKM, {\it i.e.,} the strong central attraction.

Fig. 2 and Fig. 3 show the central potential $V^{T}_{C}$ calculated
from Eq. (\ref{35}) and the first term of the right hand of Eq.
(\ref{35}) only. In order to keep the figures clear, we plot the
potential curves of  $T=0$ and $T=1$ separately.  For the case with
the nucleon only, the results of $V^{T}_{C}$ are independent of the
isospin $T$, and less attractive than the phenomenological
potentials. When the perturbation corrections due to the effects of
the intermediate states $N\overline{\Delta}$, $\Delta\overline{N}$
and $\Delta\overline{\Delta}$ (i.e., $\Delta$ mixing effects) are
taken into account, the results of $V^{T}_{C}$ show significant
attraction effects explicitly, and are closer to the Bryan-Phillips
and Nijmegen potential. These perturbation results are rather
realistic. The effects of $\Delta$($\overline{\Delta}$) mixing are
so striking in the case of $T=1$ that the perturbation result is
more attractive than the phenomenological potential for $T=1$.
Furthermore, we would like to mention that, due to isospin
conservation, the $N\Delta$ transition is missing in the $T=0$
channel which differentiates then the effect of the perturbation
result between the $T=0$ and the $T=1$ channels.
As a cross-check of our numerical calculation, we reproduce the
results of Ref. \cite{amado2}, the nucleon only results of Ref.
\cite{amado2} are also shown in Fig. 2 and Fig. 3 in order to
illustrate the role of $B^{\mu}B_{\mu}$ term. From these figures it
can be seen that the central potentials from the Mod-SKM are in
better agreement with the phenomenology potentials.

In Fig. 4 and Fig. 5, we show the $T=0$ and $T=1$ spin-dependent
potentials. In these cases the nucleon only potential and the
perturbative results are quite similar. From 1 fm to about 1.5 fm,
the potentials from the modified Skyrme model are not so close to
the phenomenological potentials. Especially in the $T=0$ case, both
the nucleon only and perturbative analysis give a positive spin-spin
potential, in contrast to the negative values of the
phenomenological potentials. It is important to see if the more
complete Skyrme calculations can repair this disagreement. However,
the smallness of the potential is reproduced. In our calculation,
that small value arises from the cancelations of large terms.

Fig. 6 and Fig. 7 show the tensor potential $V^{T}_{t}$. Being
similar to case of the spin-dependent potential,  the nucleon only
potential and the perturbative results are also quite similar.
Particularly at large distance, these results agree with the
phenomenological potential, but the agreement is not so good at
intermediate distance. However, the difference between the
theoretical and the phenomenological results is of the order of 10
MeV, compared to the static soliton mass or the nucleon mass which
is about 1GeV, the difference is small enough. Here again, an
improved Skyrme model dynamical calculation, going beyond the
product ansatz, using diagonalization for state mixing and including
explicitly the vector meson ($\rho,\omega$) and some high derivative
terms in the Lagrangian, might lead to a better agreement.

\section{conclusion and discussion}
We have shown that the modified Skyrme model with product ansatz can
give $N\overline{N}$ interaction which is in qualitatively agreement
with the phenomenological potential, and it provides a better
description of the static properties of single baryon than the
minimal version Skyrme model. We see that the configuration mixing
is very important to be included, and we roughly estimate this
effect by perturbation theory. More sophisticated method of
considering the state mixing effect is the Born-Oppenheimer
approximation. The potential curves in the Born-Oppenheimer
approximation are similar to the perturbative results, especially
for the spin-dependent and the tensor potential
\cite{nucleon5,amado2}.

To go from this work to a theory that can be confronted with
experiment in detail is a difficult challenge, {\it i.e,} predicting
the nucleon-antinucleon scattering cross section, the polarization,
and the spectrums of the nucleon-antinucleon system {\it etc}. There
are non-adiabatic effects that are particularly important at small
$R$, and there are other mesons which should be included in the
Skyrme lagrangian. The effects due to vector mesons may be
particularly important at small distances. Obtaining the static
nucleon-antinucleon interaction from Skyrme model based on large
$N_C$ QCD can be a promising approach. We expect we can further
discuss whether or not there exists nucleon-antinucleon bound
state(baryonium) in this framework.
\section *{ACKNOWLEDGEMENTS}
\indent

We would like to thank Hui-Min Li and Yu-Feng Lu for their help with
the numerical calculations. The computations were carried out at
USTC-HP Laboratory of High Peformance Computating. This work is
partially supported by National Natural Science Foundation of China
under Grant Numbers 90403021, and by the PhD Program Funds
20020358040 of the Education Ministry of China and KJCX2-SW-N10 of
the Chinese Academy.

\begin{figure}[hptb]
\begin{center}
\label{fig1}
\includegraphics*[width=11cm]{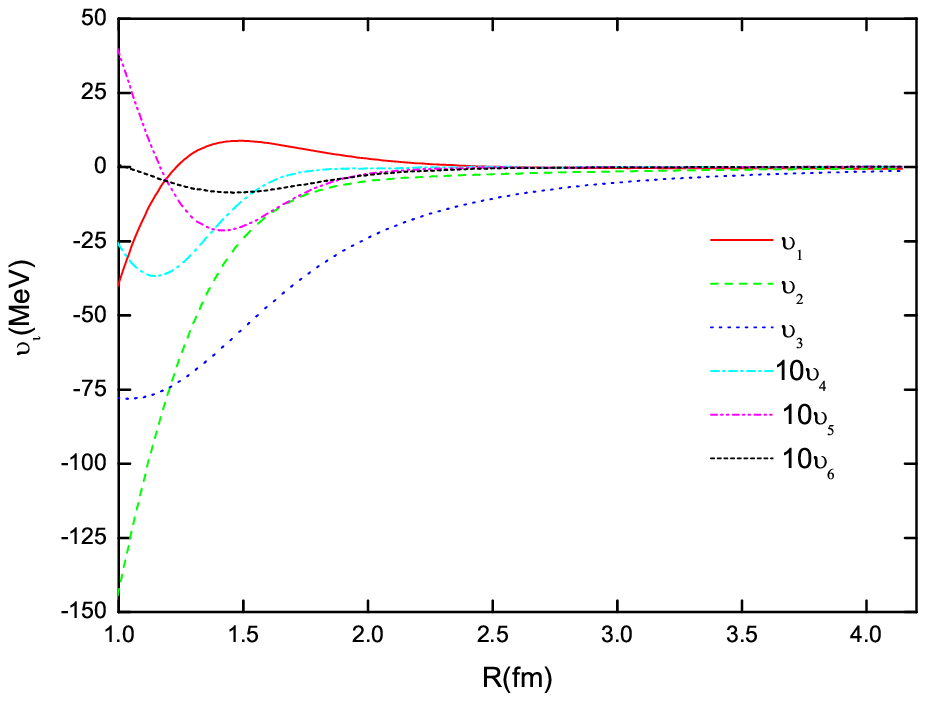}
\caption{Various terms of the Skyrmion-antiSkyrmion potential
potential Eq. (\ref{29}). $\upsilon_1(R)$ is shown by the solid
line, $\upsilon_2(R)$ by the dashed line, $\upsilon_3(R)$ by the
dotted line, $\upsilon_4(R)$ by the dash-dotted line,
$\upsilon_5(R)$ by the dash-dot-dotted line, and $\upsilon_6(R)$ by
the short dash-dotted line. }
\end{center}
\end{figure}

\begin{figure}[hptb]
\begin{center}
\label{fig2}
\includegraphics*[width=11cm]{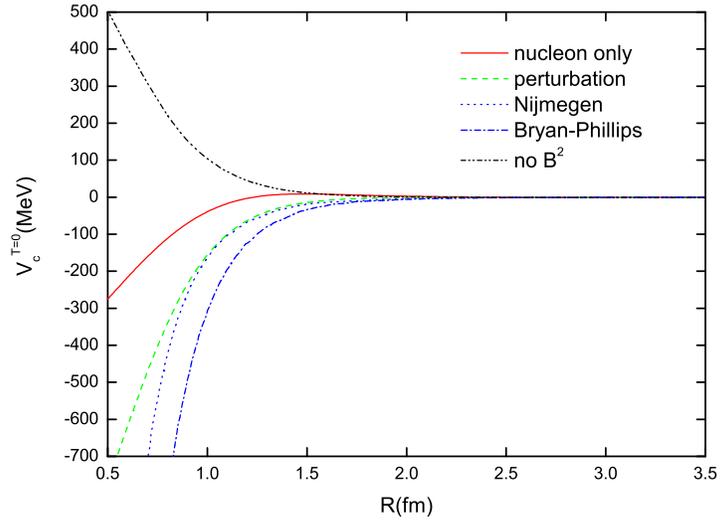}
\caption{The central potential $V^{T}_c$ as a function of distance
$R$ for the $T=0$ channel. The solid line is the nucleon only
result, the dashed line is the result of the states mixing by
perturbation theory. The dash-dot-dotted line is the nucleon only
potential in conventional Skyrme model \cite{amado2}. The
phenomenological potentials based on meson exchange are shown by
dash-dotted line for the Bryan-Phillips potential and by dotted line
for the Nijmegen potential.}
\end{center}
\end{figure}

\begin{figure}[hptb]
\begin{center}
\label{fig3}
\includegraphics*[width=11cm]{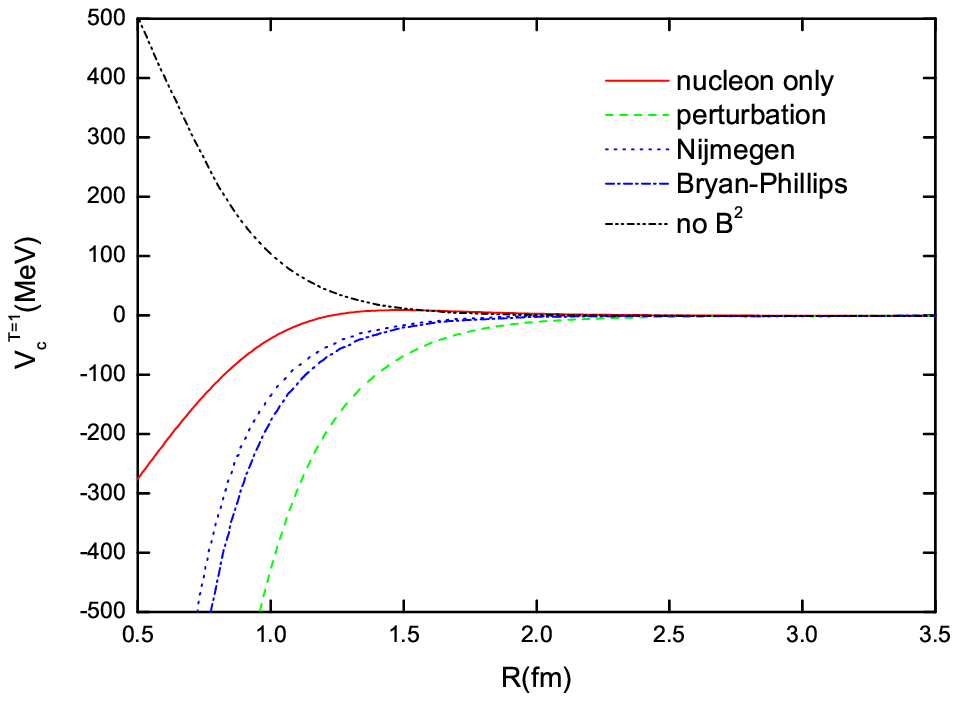}
\caption{The central potential $V^{T}_c$ as a function of distance
$R$ for the $T=1$ channel, labeling of the curves is the same as
that in Fig. 2.}
\end{center}
\end{figure}

\begin{figure}[hptb]
\begin{center}
\label{fig4}
\includegraphics*[width=11cm]{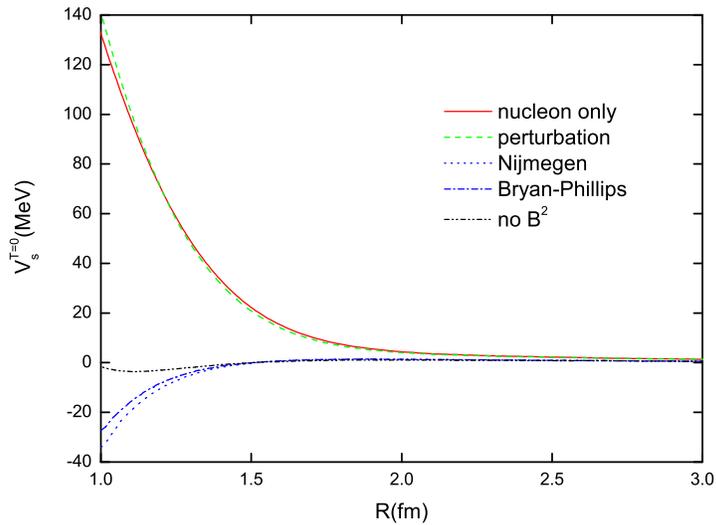}
\caption{The spin-dependent potential $V^{T}_s$ as a function of
distance $R$ for the $T=0$ channel, labeling of the curves is the
same as that in Fig. 2.}
\end{center}
\end{figure}

\begin{figure}[hptb]
\begin{center}
\label{fig5}
\includegraphics*[width=11cm]{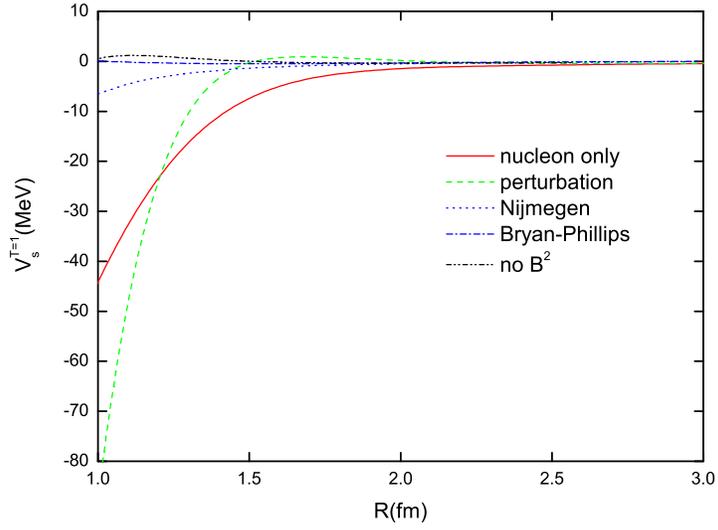}
\caption{The spin-dependent potential, same as Fig. 4 but for
$T=1$.}
\end{center}
\end{figure}

\begin{figure}[hptb]
\begin{center}
\label{fig6}
\includegraphics*[width=11cm]{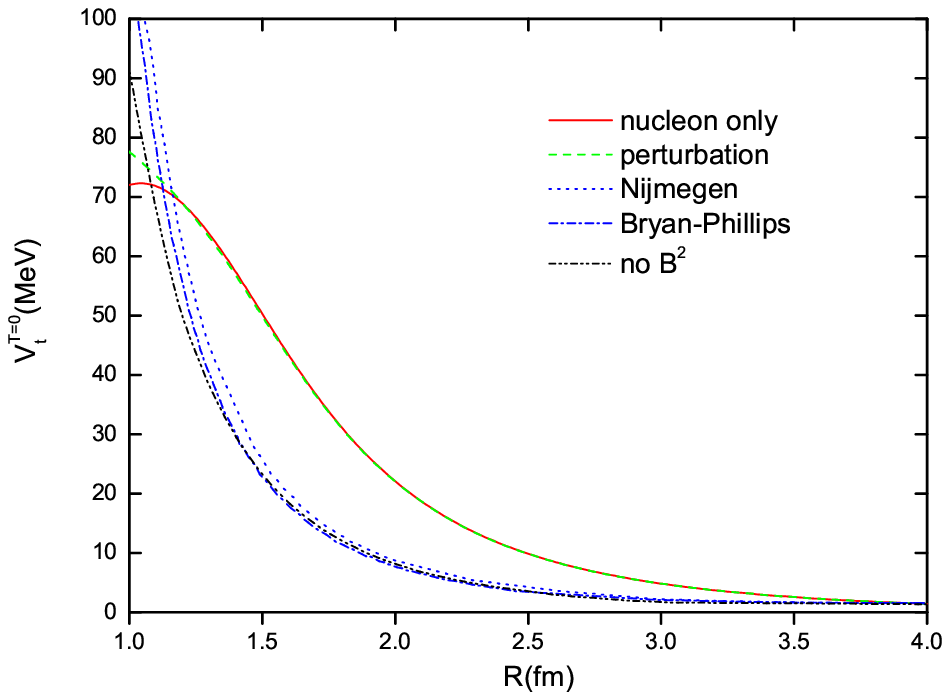}
\caption{The tensor potential $V^{T}_t$ as a function of distance
$R$ for the $T=0$ channel, labeling of the curves is the same as
that in Fig. 2.}
\end{center}
\end{figure}

\begin{figure}[hptb]
\begin{center}
\label{fig7}
\includegraphics*[width=11cm]{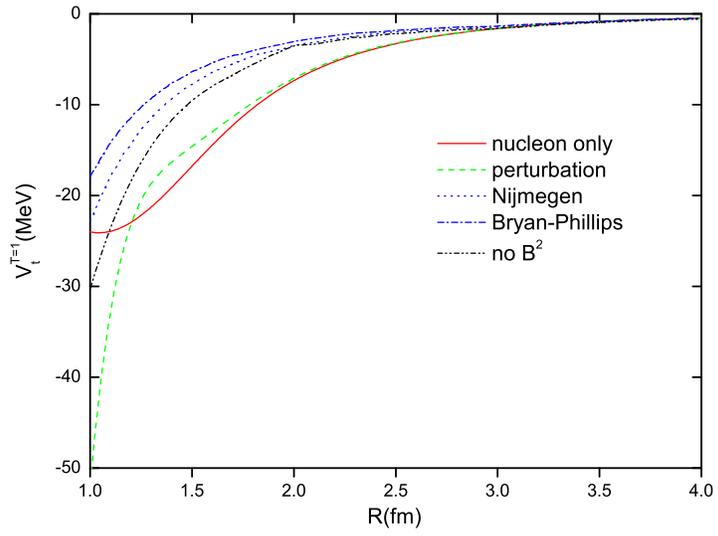}
\caption{The tensor potential, same as Fig. 6 but for $T=1$}
\end{center}
\end{figure}


\end{document}